\documentclass[apj]{emulateapj}
\usepackage{apjfonts}
\usepackage{psfrag}

\shorttitle{Spatially-Resolved H$\alpha$ Maps at $z\sim 1$}
\shortauthors{Nelson et al.}

\begin{document}

\gdef\ha{H$\alpha$}
\gdef\ew{${\rm EW}({\rm H}\alpha)$}

\title{Spatially resolved \ha\ maps and sizes of 57 strongly star-forming galaxies 
at $z\sim1$ from 3D-HST: evidence for rapid inside-out assembly of disk galaxies}
\slugcomment{Accepted for publication in ApJ Letters}

\author{Erica J.\ Nelson\altaffilmark{1}, Pieter G.\ van Dokkum\altaffilmark{1}, Gabriel
Brammer\altaffilmark{2},
Natascha F\"orster Schreiber\altaffilmark{3},
Marijn Franx\altaffilmark{4},
Mattia Fumagalli\altaffilmark{4},
Shannon Patel\altaffilmark{4},
Hans-Walter Rix\altaffilmark{5},
Rosalind E.\ Skelton\altaffilmark{1},
Rachel Bezanson\altaffilmark{1},
Elisabete Da Cunha\altaffilmark{5},
Mariska Kriek\altaffilmark{6},
Ivo Labbe\altaffilmark{4},
Britt Lundgren\altaffilmark{1},
Ryan Quadri\altaffilmark{7},
Kasper B.\ Schmidt\altaffilmark{5}}

\altaffiltext{1}{Astronomy Department, Yale University, New Haven, CT 06511}
\altaffiltext{2}{European Southern Observatory, Alonson de C\'ordova
3107, Casilla 19001, Vitacura, Santiago, Chile}
\altaffiltext{3}{Max-Planck-Institut f\"ur extraterrestrische Physik,
Giessenbachstrasse, D-85748 Garching, Germany}
\altaffiltext{4}{Leiden Observatory, Leiden University, Leiden, The
Netherlands}
\altaffiltext{5}{Max Planck Institute for Astronomy (MPIA), K\"onigstuhl 17,
69117, Heidelberg, Germany}
\altaffiltext{6}{Department of Astronomy, University of California,
Berkeley, CA 94720, USA}
\altaffiltext{7}{Carnegie Observatories, 813 Santa Barbara Street, Pasadena,
CA 91101, USA}


\begin{abstract}

We investigate the build-up of galaxies at
$z\sim1$ using maps of \ha\ and stellar continuum emission for a
sample of 57 galaxies with rest-frame \ha\ equivalent widths
 $>100$\,\AA\ in the 3D-HST grism survey. We find that the
\ha\ emission broadly follows the rest-frame $R$-band
light but that it is typically somewhat more extended and
clumpy. 
We quantify the spatial distribution with the half-light radius. 
The median \ha\ effective radius $r_e({\rm H}\alpha)$ is $4.2\pm0.1$ kpc
but the sizes span a large range,
from compact objects with $r_e({\rm H}\alpha) \sim1.0$\,kpc to extended 
disks with $r_e({\rm H}\alpha)\sim15$\,kpc. Comparing \ha\ sizes to 
continuum sizes, we find $\left<r_e({\rm H}\alpha)/r_e(R)\right>=1.3\pm{}0.1$ 
for the full sample. 
That is, star formation, as traced by H$\alpha$, typically occurs out to larger
 radii than the rest-frame R-band stellar continuum; galaxies are growing their radii and building up from the inside out. 
This effect appears to be somewhat more pronounced for the largest galaxies.
Using the measured \ha\ sizes, we derive star formation rate
surface densities, $\Sigma_{\rm{}SFR}$. We find that $\Sigma_{\rm{}SFR}$
ranges from $\sim{}0.05$\,M$_{\odot}$\,yr$^{-1}$\,kpc$^{-2}$
for the largest galaxies to $\sim 5$\,M$_{\odot}$\,yr$^{-1}$\,kpc$^{-2}$
for the smallest galaxies, implying a large range in physical conditions
in rapidly star-forming $z\sim{}1$ galaxies. Finally, 
we infer that all galaxies in the sample have very high gas mass fractions and
stellar mass doubling times $<500$\,Myr. 
Although other explanations are also possible,
a straightforward interpretation is
that we are simultaneously witnessing
the rapid formation of compact bulges and large disks 
at z$\sim{}1$.

\end{abstract}

.

\section{Introduction}

Many galaxies in the Universe today are in their twilight years,
with low star formation rates compared
to the already assembled stellar mass. As an example,
the specific star formation rate
(SSFR, the star formation rate divided by the stellar mass) of our own
Galaxy
is $\sim 5\times 10^{-2}$\,Gyr$^{-1}$, which means that at its
present growth rate it would take the Milky
Way $\sim 20$\,Gyr to double its stellar mass.
By contrast, 
at $z\gtrsim 1$ global star formation rates were much
higher than they are today (e.g., Lilly {et~al.} 1996; Madau {et~al.} 1996; {Geach} {et~al.} 2008)
 and it was probably during this epoch that $\sim L_*$
galaxies assembled the bulk of their stellar mass. 

Recent studies have shown that the star formation rate of distant galaxies
is correlated with their stellar mass, and that  the relation of star formation rate with mass evolves with redshift (e.g., {Zheng} {et~al.} 2007; {Damen} {et~al.} 2009; {Noeske} {et~al.} 2007; {Elbaz} {et~al.} 2007; {Daddi} {et~al.} 2007; Karim {et~al.} 2011). 
Using high spatial resolution
imaging, it has also been shown that
star formation rates correlate with the
structural properties of the existing stellar populations:
galaxies that are actively forming stars typically have low Sersic index
and surface density, whereas quiescent galaxies have high Sersic
index and surface density (e.g., {Kauffmann} {et~al.} 2003; {Franx} {et~al.} 2008; {Wuyts} {et~al.} 2011; {van Dokkum} {et~al.} 2011; {Bell} {et~al.} 2012). 

These global relations tell us {\em which} galaxies are growing
at a particular cosmic epoch, but not {\em where} they are growing, that
is, which structural components are being built up.
In order to address the where,
we need to know the spatial distributions of
both the existing stellar population (representing a galaxy's history)
and of the ongoing star formation (representing a galaxy's future).
Star formation can be traced by the \ha\ line
emission, as it scales with the quantity of ionizing photons produced by
hot young stars ({Kennicutt} 1998). The 
stellar mass of a galaxy can be traced using the rest-frame optical
or near-IR stellar continuum emission.

Obtaining \ha\ and stellar continuum maps of distant galaxies
with $\sim 1$\,kpc resolution is challenging, and has so far only been
possible with integral field units (IFUs) on 8--10m class telescopes
(e.g., {Genzel} {et~al.} 2008; {Law} {et~al.} 2009; Wright {et~al.} 2009; Wisnioski {et~al.} 2011; Stark {et~al.} 2008; Jones {et~al.} 2010; {Yuan} {et~al.} 2011; {Kriek} {et~al.} 2009). 
These studies have shown
that many star forming galaxies at $z\sim 2$
are clumpy and irregular in their $H\alpha$ emission.
This is also seen in the rest-frame UV (e.g., Lotz {et~al.} 2006; Ravindranath {et~al.} 2006).
For a small sample of $z\sim 2$
galaxies observed with the NICMOS camera on the Hubble Space Telescope
(HST), {F{\"o}rster Schreiber} {et~al.} (2011a, 2011b)
showed that the spatial extent of their stars and star formation was
very similar, even though individual clumps of star formation
do not necessarily coincide with clumps of already-existing stars.

Here we build on these studies by analyzing a sample of 57 galaxies
with high resolution \ha\ and stellar continuum maps at $z\sim 1$,
a time during which galaxies probably
transitioned from the clumpy,
irregular morphologies commonly seen at $z\sim2$ to the regular morphologies
of normal galaxies at $z\sim0$. This study has
become possible owing to 
the near-IR imaging and slitless
spectroscopic capabilities provided by the new WFC3 camera on HST.
As shown in this {\em Letter}, WFC3 can provide
maps of emission lines and continuum emission at 
$\sim 0\farcs 13$ spatial resolution. These data are
complementary to ground-based IFU studies, as they provide continuum and
line maps with a stable, high Strehl ratio
PSF at the expense of having very low spectral resolution.
We use these data, taken as part of the 3D-HST survey, to determine where
 and with what intensity galaxies at $z\sim1$ are building up their stellar mass.
We assume H$_0=70$ km s$^{-1}$Mpc$^{-1}$, $\Omega_M=0.3$, 
and $\Omega_{\Lambda}=0.7$.

\section{Data}
\subsection{Sample}

H$\alpha$ maps are created from the spatially resolved spectra of
3D-HST, a 248-orbit Treasury program on the Hubble Space Telescope
(Brammer et al. 2012,  {van Dokkum} {et~al.} 2011). Together with
data taken in a previous program, 3D-HST will cover the
five well-studied extra-galactic fields that are also covered by
the CANDELS imaging program ({Grogin} {et~al.} 2011; {Koekemoer} {et~al.} 2011):
AEGIS, COSMOS, GOODS-North \footnote{Data taken in program GO-11600 (PI: B. Weiner).}, 
GOODS-South, and UDS.
The data presented in this paper are based on the $\sim70$ pointings,
roughly half of the full data set, that were obtained prior to June
2011 (see  {van Dokkum} {et~al.} 2011,  Brammer et al. 2012).


The WFC3 G141 grism provides spatially resolved spectra of all
sources in the field.  The wavelength
range of the G141 grism,
$1.15\mu{\rm m} < \lambda < 1.65\mu$m, covers the
H$\alpha$ emission line for galaxies in the redshift range
$0.7<z<1.5$. The survey also provides
broad-band near-infrared imaging in the F140W filter, which samples
the rest-frame $R$ band for $z\sim 1$.
High spatial resolution maps of $H\alpha$
and continuum emission taken with the same camera under the same
conditions is a key strength of this program as it allows a direct
comparison of the structural features present in the distribution of the light of 
stars and of star formation. The spatial resolution is $\approx 0\farcs 13$,
sampled with $0\farcs 06$ pixels,
corresponding to 0.5\,kpc at $z\sim1$.

The data reduction uses the aXe code ({K{\"u}mmel} {et~al.} 2009) with modifications,
and is described in Brammer et al. (2012).
Redshifts and $H\alpha$ equivalent
widths (EWs) were determined from summed, one-dimensional spectra, as described
in detail in Brammer et al. (2012). 
Along with these measurements, the software
provides drizzled two-dimensional images of the galaxies in the F140W
band and drizzled two-dimensional spectra.


\begin{figure}
\centering 
\includegraphics{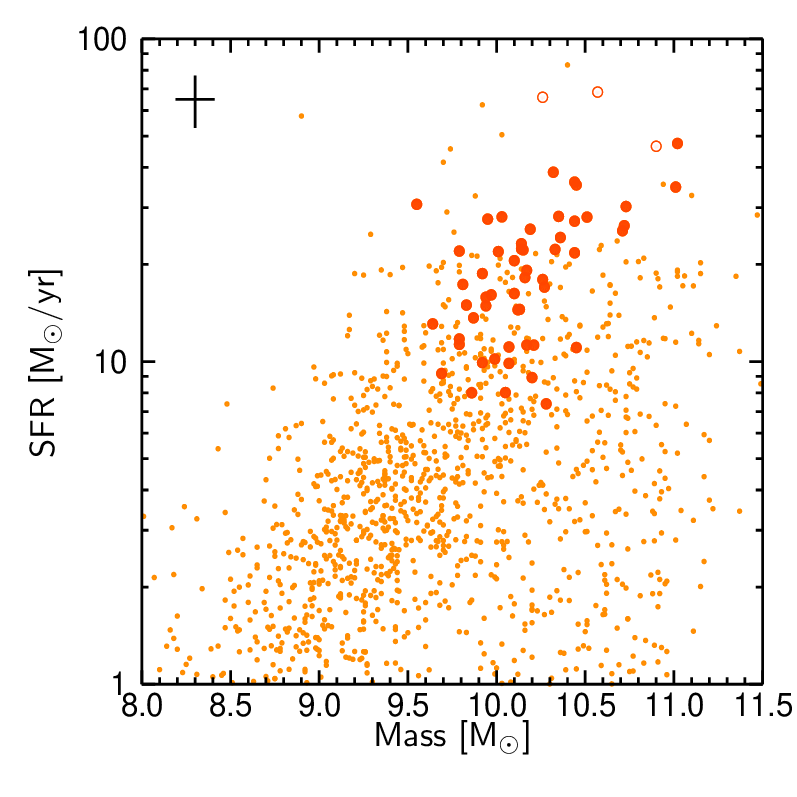} 
\caption{Distribution of the sample galaxies in the SFR-mass plane 
at $0.8<z<1.3$. No extinction correction is applied to the \ha\,\-derived SFRs.
In this study we consider galaxies with \ew\,$>100$\,\AA\,\
(the large, red circles). The errorbars shown here and in all subsequent figures
denote the typical uncertainties. 
\label{sel}}
\end{figure}

Targets were selected to have redshifts $0.8<z<1.3$,
total F140W magnitude $<21.9$ and
rest-frame \ew\,$>100$\,\AA; we thus focus
here on the objects
that are among the galaxies most  actively forming stars relative to their
existing stellar mass, during a transitional epoch in which galaxies were
assembling the structure we see today. This is illustrated in Fig.\ \ref{sel},
which shows the distribution of the selected galaxies in the SFR--mass
diagram (see e.g., {Wuyts} {et~al.} 2011).
Stellar masses	were derived using the FAST code ({Kriek} 
{et~al.} 2009a) (see	{Brammer} {et~al.} 2012) and star formation 
rates were determined from 
the \ha\ luminosities using the {Kennicutt} (1998) relation. 
  These criteria further ensure reliable and high S/N
measurements of the line emission.  Due to the nature of slitless
grism spectroscopy, a spectrum can be contaminated by flux from
overlapping spectra from the object's neighbors. 
As a final criterion we limited the sample to objects for which the
estimated contamination is less than 5\,\% of the measured flux.

The final sample comprises 57 galaxies.
We note that no selection on environment was made.
Three of the galaxies in this sample have X-ray fluxes
$L_x>5\times{}10^{43}$\,\ erg/s 
which means that they almost certainly host AGN. We did not remove
these galaxies from plots but marked them by open symbols. 
As we show later in Fig.\ \ref{moneyplot}, these objects have unique morphological signatures in the 2D spectra as their \ha\ line widths exceed the pixel size of $\sim1000$\,\ km/s.

\subsection{Maps of H$\alpha$ and Continuum Emission}

\begin{figure*}
\centering
\includegraphics[width=1.02\textwidth]{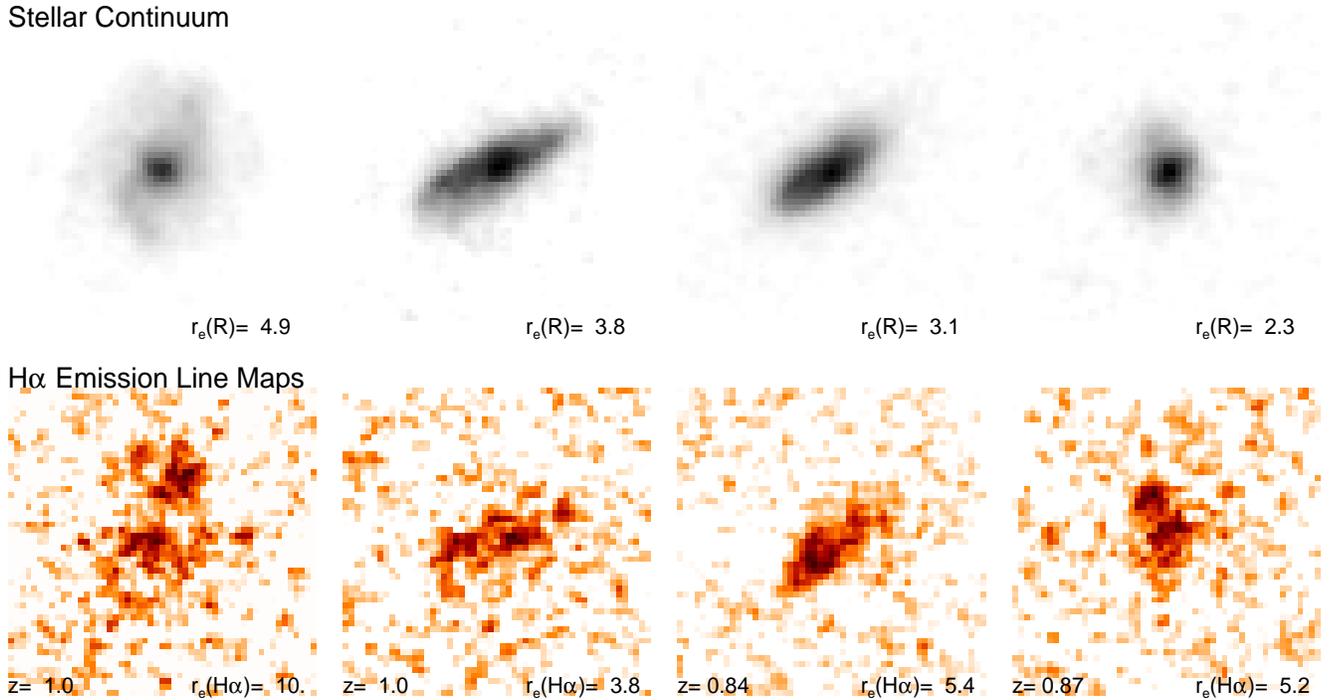} 
\caption{Examples of galaxies in the survey. Each panel shows the
continuum emission, as traced by the F140W filter (rest-frame $R$),
and the H$\alpha$ emission line map.
\label{exmaps}}
\end{figure*}

The combination of the WFC3 camera's high native spatial resolution
and the grism's $R\sim130$ point source spectral resolution
yields a spectrum that is best envisaged as images of a galaxy taken
at $46$\,\AA\,\  increments (23\,\AA\,\ after drizzling) and placed next to each other on the
detector. Thus, an emission feature in a high spatial resolution
slitless spectrum is essentially an image of a galaxy at that wavelength.
To create an emission line map of a galaxy, we masked the emission
line and fit a polynomial to the continuum of its collapsed,
one-dimensional spectrum. This model fit was then scaled and
subtracted row-by-row from the two-dimensional spectrum; the residual
is a map of the galaxy in the light of the $H\alpha$ line.

Fig.\ \ref{exmaps} shows
emission line maps and their corresponding F140W (rest-frame optical)
images for eight representative galaxies which illustrate
trends seen in the full sample.
Inspecting the \ha\ maps and the continuum maps,
we find that the spatial distribution of \ha\
broadly follows the spatial distribution of stellar light.
However, as has been seen previously at higher redshift,
H$\alpha$ often appears to be clumpier and
more asymmetric than its stellar counterpart ({Genzel} {et~al.} 2008; {Law} {et~al.} 2009, 2011; {F{\"o}rster Schreiber} {et~al.} 2011a, 2011b). 
 In the next section we quantify the spatial extent of the $H\alpha$ and
stellar continuum emission by measuring their sizes. 

Several uncertainties in the interpretation of the line maps should
be emphasized here. First,
\ha\ $\lambda6563$ and [N\,{\sc ii}]
$\lambda6583$ are not resolved with the G141 grism, and \ha\ and 
[S\,{\sc ii}] $\lambda\lambda 6716,6731$ are barely resolved.
The unknown ratios of these lines lead to uncertainties in
\ew, but also in the H$\alpha$ morphologies, as the line ratios
almost certainly vary with position ({Genzel} {et~al.} 2008;
Cresci {et~al.} 2010; {Yuan} {et~al.} 2011; {Queyrel} {et~al.} 2011).
A second systematic uncertainty is dust attenuation. 
Differential extinction
could affect the spatial distribution in the maps and the
resulting size measurements. 
Third, the position of the
H$\alpha$ emission relative to the continuum emission in
the wavelength (``$x$'') direction is degenerate
with the redshift of the object. 
\footnote{Note that this does not affect our size measurements in \S3.}
Finally, the F140W filter contains the H$\alpha$ line, introducing
a built-in correlation between the emission line maps and the
continuum maps. This effect is small, as
for galaxies in our sample the $H\alpha$ line
typically contributes only $\sim 5\%$ to the F140W flux.

\section{H$\alpha$ and continuum Sizes}

\begin{figure*}
\centering
\includegraphics[width=\textwidth]{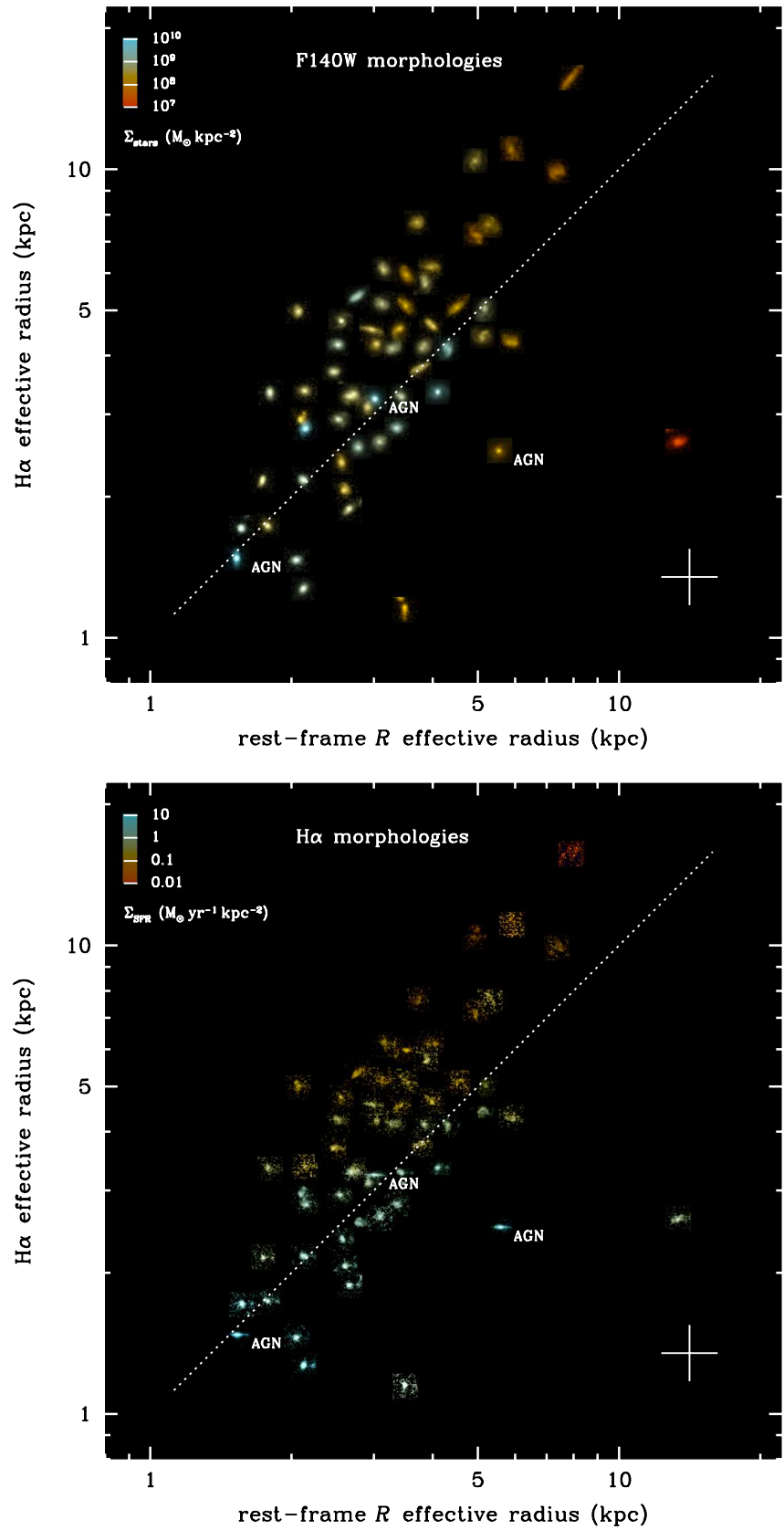}
\caption{Relation between half-light radius of the \ha\ emission and the
half-light radius of the stellar continuum radiation (as described by the
F140W imaging). The images show the morphology in F140W (top panel)
and in \ha\ (bottom panel). 
The \ha\ size scales with the rest-frame optical size but is typically somewhat larger.
\label{moneyplot}}
\end{figure*}

A first order parameterization of the \ha\ distribution is its effective radius
(or half-light radius).
The standard method for determining sizes of distant galaxies is to
fit two-dimensional {Sersic} (1968) models, convolved with the point spread
function (e.g., Peng {et~al.} 2010). However, such
parametric methods do not provide meaningful results for the
\ha\ images, as they are typically clumpy and often have asymmetrical or
centrally-depressed light profiles (Shapley 2011).
 Instead, we measured effective radii (half-light radii) for both the \ha\ images
and the rest-frame
optical continuum images as follows. We first deconvolved the images, using
empirical PSFs and a variation of the CLEAN algorithm. 
We verified that the key results do not depend on the deconvolution algorithm.
Next, growth curves were calculated using circular apertures centered
on the luminosity-weighted mean $x$ and $y$ positions. The \ha\ center
was not forced to coincide with the F140W center. The effective radii were
defined as the radii containing 50\,\% of the maximum growth curve value.
From varying the deconvolution algorithm and the sky determination 
we estimate typical errors of $\sim 0.06$ dex. 

We find that the \ha\ sizes span a very large range, from 1\,kpc to
15\,kpc. The immediate implication is that star formation
in strongly star-forming galaxies
at $z\sim1$ occurs on a large range of physical scales, from the
sizes of compact objects (such as bulges) to those of large disks.
In Fig.\ \ref{moneyplot} we compare the H$\alpha$
sizes to the rest-frame optical sizes of the galaxies.
This comparison provides information on where galaxies are
growing: galaxies for which the effective
radius of star formation 
is equal to their stellar continuum radius
should experience little structural change, as their current
growth follows their past growth. Galaxies for which
$r_e({\rm H}\alpha)>r_e(R)$ have
star formation
which is more extended than their assembled stellar mass, implying
that their growth began with a smaller central region and has
proceeded outward: inside-out growth. This might be expected if
gas is accreted onto a galaxy and then
cools onto the galaxy's disk (e.g., Brooks {et~al.} 2009).
Galaxies that have $r_e({\rm H}\alpha)<r_e(R)$ have
star formation which is more concentrated than their stars, which
might happen
after a gas-rich merger producing a central star burst.
Thus, the location of a galaxy on a \ha\ size vs.\ rest-frame optical size
diagram provides information about where it's growing.

As is clear from Fig.\ \ref{moneyplot} the $H\alpha$ sizes
generally track the F140W (rest-frame $R$ band) sizes but are typically
somewhat larger.
Excluding the three AGN, the median ratio of star formation size to rest-frame
optical size for our sample is $r_e({\rm H}\alpha)/r_e(R)=1.3\pm{}0.1$,
where the $1\sigma$ error was derived from bootstrap
resampling. The ratio has an rms scatter of $\pm 0.2$\,dex,
which implies that there
is a considerable range in the spatial properties of growth
for the galaxies in this sample. 

This scatter reflects partly the fact that for a given stellar continuum size
there is a range in \ha\ sizes.
It also reflects the
fact that the relation between \ha\ size and rest-frame $R$ band size
appears slightly steeper than linear. Dividing the galaxies into two
subsamples based
on their \ha\ sizes 
shows that the smallest 50\,\% of galaxies have
median $r_e({\rm H}\alpha)/r_e(R)=1.0\pm{}0.1$,
while the largest 50\,\% have median $r_e({\rm H}\alpha)/r_e(R)=1.5\pm{}0.1$.
That is, galaxies with extended $H\alpha$
emission in an absolute sense also tend to have $H\alpha$ which is
more extended relative to their already-constructed stellar component.
The range of values of the \ha\ to stellar continuum radius suggest
that we may be observing the construction of different structural
components in different galaxies: from a central concentration to an
extended disk around a previously formed central concentration.


\section{Star Formation Surface Densities}


\begin{figure}
\centering
\includegraphics{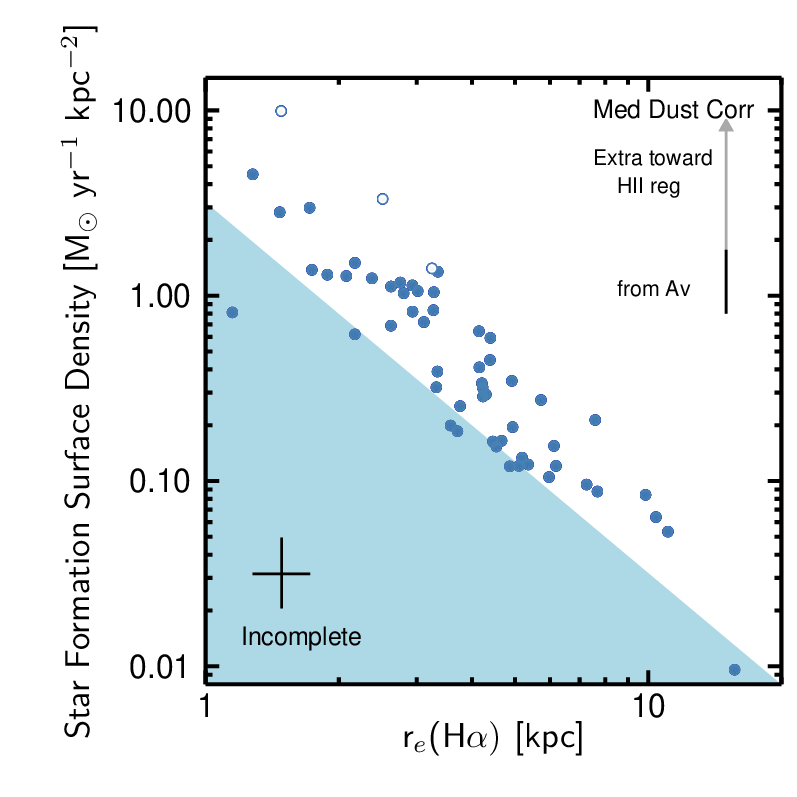}
\caption{
Star formation surface density as a function of \ha\ size of galaxies.
 The lowest star formation rates we detect 
in this sample are $\sim{}10{}M_{\odot}/yr$ (Fig.\ 1) and the
 blue region indicates the 
surface density incompleteness implied by this limit. 
The points shown are uncorrected for dust attenuation and the 
arrow shows the estimated dust correction. 
The median extinction derived from SED fitting to the continuum is black and the 
additional attenuation estimated to exist toward
  HII regions is gray ({Calzetti} {et~al.} (2000)).
  The star formation surface densities span a range of a factor of
  $\sim100$.
\label{sSFRSD}
}
\end{figure}

The star formation surface density, i.e., the star formation
rate per unit area, is thought to be a key
parameter in probing the physics 
driving the conversion of gas to stars
 ({Genzel} {et~al.} 2010; {Daddi} {et~al.} 2010; {Wuyts} {et~al.} 2011; {Rodighiero} {et~al.} 2011).
Star formation surface densities are generally determined from
stellar continuum sizes,
under the assumption
that star formation occurs over the same area as the emitted
stellar continuum light.
Here, however, we have measured $r_e$(\ha\ ), the radius over which star 
 formation is occurring, and we can directly determine

$$\Sigma({\rm SFR})=SFR/\pi{}r_e({\rm H}\alpha)^2$$
i.e., the star formation rate divided by the area covered by the
line-emitting gas. 
The scatter in Fig.\ \ref{moneyplot} implies that
star formation surface densities determined
from continuum radii (as done in e.g., {Wuyts} {et~al.} 2011)
have typical errors of $\sim 0.4$ dex.

In Fig.\ \ref{sSFRSD} we show $\Sigma$(SFR) as a function of \ha\ size. 
It is clear that there is a very large range. Small galaxies have much 
higher star formation densities than large galaxies
 but that is mostly driven by our sample selection.
The compact objects have $\Sigma_{\rm SFR} \sim
5$\,M$_{\odot}$\,yr$^{-1}$\,kpc$^{-2}$
and the most extended objects have $\Sigma_{\rm SFR} \sim
0.05$\,M$_{\odot}$\,yr$^{-1}$\,kpc$^{-2}$, 
implying  the simultaneous build-up of both extended and compact 
features with high and low star formation surface densities.



\begin{figure}[htbp]
\centering
\includegraphics{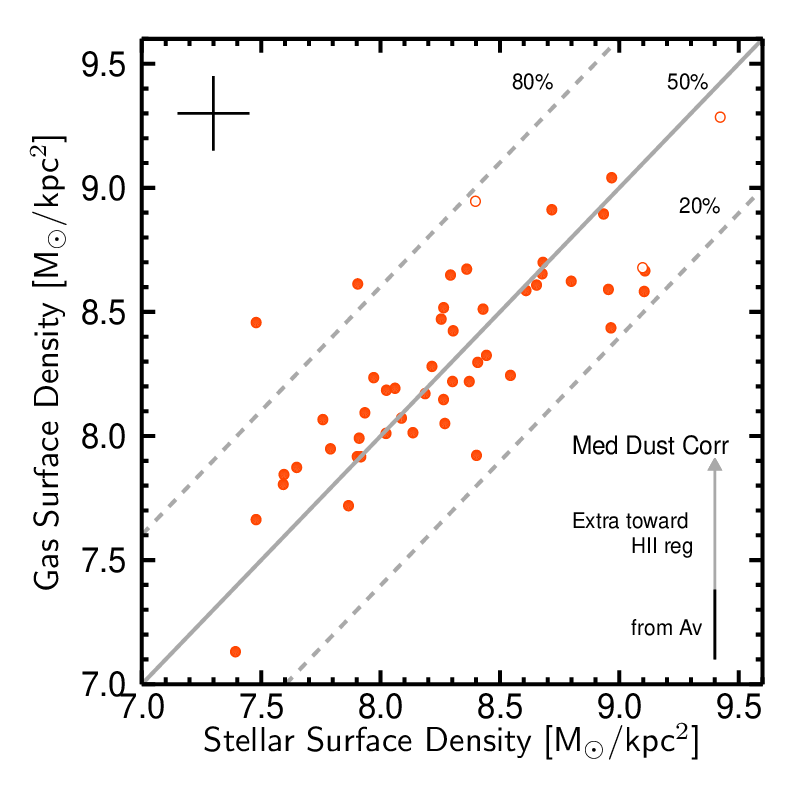} 
\caption{Gas surface density, as derived from the Kennicutt-Schmidt
law ({Kennicutt} 1998), versus stellar surface density within r$_e(H\alpha)$.
 These galaxies have high
gas mass fractions -- lines of 20\%, 50\%, and 80\% gas fraction are shown.
 The arrow shows the estimated dust correction
as described in Fig.\ \ref{sSFRSD}. 
The error bar indicates the typical error.
\label{sdgas}}
\end{figure}

The star formation surface densities (see Fig.\ \ref{sSFRSD})
of these galaxies can be used to infer 
gas surface densities using the {Kennicutt} (1998) relation. Comparing their 
inferred gas surface densities to their stellar surface densities
(Fig.\ \ref{sdgas}) suggests that most galaxies in this sample have high 
gas mass fractions $\sim50\%$, even without dust correction. 
The median dust correction as derived from SED fitting is shown
 in black and the additional attenuation estimated to exist toward
  HII regions is shown in gray ({Calzetti} {et~al.} 2000). Given the uncertainty 
 introduced by the magnitude of this possible dust correction, it will be 
 interesting to measure these gas densities directly
 (e.g., {Tacconi} {et~al.} 2010; {Daddi} {et~al.} 2010).
The high gas fractions imply the presence of large masses of gas to convert 
to stars. This conversion is occurring rapidly; the stellar
masses and star formation rates of the galaxies imply typical doubling
 times of $\sim{}500$\,Myr assuming no dust correction, or $\sim{}50$ Myr using the maximum dust correction.



\vspace{5 mm}
\section{Discussion}

Using new data from the 3D-HST survey, we have constructed
high-resolution maps of stellar continuum and
\ha\ emission to  uncover where galaxies are growing and
at what intensities. We find that the \ha\ emission generally tracks 
optical light but is slightly more extended.
The immediate implication is that the galaxies have gradients in
their stellar population properties, e.g., in $A_V$, metallicity, or age. We cannot reliably
determine the cause of the gradient without additional data, particularly on the dust distributions.
However, the most straightforward interpretation is that the youngest stars have a more
extended distribution than older stars, and that the galaxies grow inside-out.
This result
was expected from the observed size evolution of star forming galaxies
 (e.g., {Williams} {et~al.} 2010)
 and consistent with theoretical expectations 
({Sales} {et~al.} 2009; {Dutton} \& {van den Bosch} 2011).
 Interestingly, this size difference has not been seen in
IFU observations of $z\sim{}2$ galaxies ({F{\"o}rster Schreiber} {et~al.} 2011a). 
More compact galaxies, on the other hand, do not appear to be in 
the process of growing their radii with a median $r_e(H\alpha)/r_e(R)$ of one.

The star formation surface densities of the galaxies, representing
the intensity of star formation, show a range of two orders of
magnitude. All galaxies have very high gas fractions, $\gtrsim 50\%$.
Thus, we see the build-up of very compact and very extended components 
with high gas fractions, implying that these components are in the early 
phases of assembly. It is tempting to suggest that the galaxies with extended, 
low surface density star formation are in the process
of building their disks while galaxies with 
compact, high surface density star formation are building bulges or
compact early-type galaxies.


It is not clear how the compact galaxies and the large galaxies
are related. A possible interpretation for the large range in
properties is that we are looking at an evolutionary
sequence in which the compact, rapidly star-forming galaxies are the
young stages of the extended galaxies. For instance, galaxies could
rapidly build up their bulges through compact star formation and then
subsequently more slowly grow their extended disks by continuous gas
accretion. However, the reverse may also be possible
(e.g. {Dekel} {et~al.} 2009, {Hopkins}{et~al.} 2011). It may also be that some of the compact, high-intensity star-forming
galaxies will evolve into compact quiescent galaxies similar to 
those seen frequently at higher redshifts
(e.g., {van Dokkum} {et~al.} 2008). In this context
it is interesting to note that the total baryonic surface
densities of the densest objects ($\sim 10^9 M_\odot/\textrm{kpc}^2$) approach the
densities of early-type galaxies at $z\sim1$ ({Franx} {et~al.} 2008).

This study can be expanded upon in many ways.
The galaxies selected for this
paper represent the most actively star-forming galaxies at the peak
epoch of global star formation. In future papers we intend to 
include more typical star-forming galaxies at this epoch. It
will also be interesting to extend this study to both higher and
lower redshifts: it may be that the properties of galaxies are
very similar at fixed \ew, independent of redshift.
As noted in \S2, both dust extinction and the contribution of [N\,{\sc{}ii}] and
[S\,{\sc{}ii}] to the \ha\
line emission are fundamental uncertainties in both the quantities
measured and spatial distribution of \ha\ flux. Assessing the
contribution of [NII] requires mapping on a spectrograph with higher
spectral resolution such that \ha\ and [NII] are
distinguishable. 
Differential dust extinction could conceivably be
corrected using spatially-resolved maps of Balmer decrements.




\end{document}